\documentstyle[12pt]{article}
\advance\textheight by \topskip
\begin{document}
\thispagestyle{empty}
\begin{flushright}
SU--ITP--94--24\\
IEM--FT--88/94\\
hep-th/9408023\\
\today
\end{flushright}
\vskip 1cm
\begin{center}
{\Large\bf STATIONARITY OF INFLATION}
\vskip 0.5cm
{\Large\bf  AND PREDICTIONS OF QUANTUM COSMOLOGY}
\vskip 1.4cm
{\bf Juan Garc\'{\i}a--Bellido}\footnote{
E-mail: bellido@slacvm.slac.stanford.edu}
\ \ and \ \
{\bf Andrei Linde}\footnote{E-mail: linde@physics.stanford.edu}
\vskip 1.5mm
Department of Physics, Stanford University, \\
Stanford, CA 94305-4060, USA
\end{center}

\vskip 1cm

{\centerline{\large\bf Abstract}}
\begin{quotation}
\vskip -0.4cm
We describe several different regimes which are possible in
inflationary cosmology. The simplest one is inflation without
self-reproduction of the universe. In this scenario the universe
is not stationary. The second regime, which exists in a broad
class of inflationary models, is eternal inflation with the
self-reproduction of inflationary domains. In this regime local
properties of domains with a given  density and given values of
fields do not depend on the time when these domains were
produced. The probability distribution to find a domain with
given properties in a self-reproducing universe may or may not
be stationary, depending on the choice of an inflationary model.
We give  examples of  models where  each of these possibilities
can be realized, and discuss  some implications of our results
for quantum cosmology.  In particular, we propose a new
mechanism which may help  solving the cosmological constant
problem.

\end{quotation}
\newpage

\section{Introduction}
According to the first versions of inflationary theory,
inflation was an important but extremely short intermediate
stage of the evolution of the Universe. Later it was discovered
that in many versions of this theory
inflation never ends because of the process of self-reproduction
of inflationary domains. This realization dramatically changed
our point of view on the fate of the Universe and on its global
structure  \cite{MyBook}.

Self-reproduction of the inflationary universe is possible both
in the old inflationary scenario \cite{b51}, and in the new
inflationary scenario \cite{b52,b62}.  However, the significance
of the existence of this regime was not fully realized until it
was shown to occur in the chaotic inflation scenario
\cite{b19}. In the simplest versions of this scenario
inflationary domains may jump for indefinitely long time at
densities close to the Planck density \cite{LLM}. Quantum fluctuations
of all physical fields and metric are extremely large in such
domains. As a result, the Universe becomes divided into
exponentially large domains filled with matter with all possible
types of symmetry breaking \cite{MyBook}, and maybe even with
different types of compactification of space-time \cite{Zeln}.
Variations in the laws of low-energy physics in different domains
are typically discrete, such as the change of
symmetry breaking from $SU(5)$ to $SU(3)\times SU(2)\times
U(1)$.  However, continuous changes are also possible, such as
the change of an effective gravitational constant in the
inflationary Brans-Dicke cosmology \cite{GBLL}.

This gave us a possibility to justify the weak anthropic
principle in the context of inflationary cosmology, and even to
speculate about the Darwinian approach to particle physics and
cosmology \cite{PhysToday} and about some kind of natural
selection of the ``constants'' of particle physics which lead to
a greater physical volume of those domains which can be occupied
by observers of our type \cite{MyBook,LLM,GBLL}.

The next step towards a justification of the anthropic principle in
quantum cosmology was made in \cite{CosmConst}, after
the appearance of the baby universe theory \cite{Coleman}.
This theory was based on two basic assumptions. The first
assumption is   that the coupling constants may take different
values in different quantum states of the universe. This idea is
very intriguing and it may have a good chance of being correct.
The second (and quite independent)  assumption was a particular
choice of measure on the space of all quantum states of the
universe. The choice advocated in \cite{Coleman} effectively was
based on the exponentiation of the square of the Hartle-Hawking
wave function of the universe $\Psi \sim \exp\Bigl({3M_{\rm
p}^4\over 16 V}\Bigr)$, where $V$ is the vacuum energy (or the
effective potential of the scalar field $\phi$)
\cite{OldVil,HH}.  This gave the probability to live in the
universe with the non-negative cosmological constant $\Lambda =
{8\pi\over M_{\rm p}^2} V$: \, $P(\Lambda) \sim \exp \Bigl(\exp
{3\pi M_{\rm p}^2 \over\Lambda}\Bigl)$.  This probability
distribution is peaked at $\Lambda = 0$.  However, this result
is a consequence of the ``wrong'' (negative) sign of the
gravitational action, which makes calculations unreliable.
Moreover, an extended version of this approach
suggests that the probability to obtain negative cosmological
constant is even higher \cite{Rub}.

Another possibility would be to exponentiate the square of  the
wave function   $\Psi \sim \exp\Bigl(-{3M_{\rm p}^4\over 16
V}\Bigr)$, which was first suggested by one of the present
authors \cite{Creation}, see also \cite{Creation2}. However, as
was argued in \cite{b19}, both wave functions are related to the
probability $P_c$ for some events to happen at a given point
\cite{Star}, without taking into account that  different parts
of the universe with different values of $V$ grow at a different
rate.  It may be natural therefore to use the measure of
probability $P_p$ introduced in \cite{b19}, which is
proportional to the physical volume of those parts of the
universe where such events may happen. This, combined with the
first assumption of the baby universe theory, may give us a
possibility to justify not only the weak anthropic principle,
but the strong anthropic principle as well
\cite{MyBook,CosmConst}. Moreover, the use of the probability
distribution $P_p$ can make anthropic considerations much more
precise and quantitative.

Various properties of the probability distribution $P_p(\phi,t)$
to find a given field $\phi$ in a unit physical volume have been
investigated in \cite{b19,LLM,Nambu}.  It was
found that this distribution has an important advantage over
other probability distributions.  In a sufficiently big universe
the normalized probability distribution $P_p(\phi, t)$ in many
realistic theories very rapidly approaches a stationary regime
$P_p(\phi)$, which does not depend at all on the unsettled
issues related to the probability of quantum creation of the
universe and on the choice between the Hartle-Hawking wave
function $\Psi \sim \exp\Bigl({3M_{\rm p}^4\over 16 V}\Bigr)$
and the tunneling wave function $\Psi \sim
\exp\Bigl(-{3M_{\rm p}^4\over 16 V}\Bigr)$ \cite{LLM,Mijic}.

This encouraging result indicates that the stationary
distribution $P_p(\phi, t)$ plays a very important role in
quantum cosmology. However, this approach has its own problems.
For example, as   it was shown in \cite{LLM,GBLL},   the shape
of the probability distribution $P_p$ may depend on the choice
of the time parametrization. The reason can be understood as
follows. Let us consider two {\it infinite} boxes, one with
apples, another with oranges. One can pick up one apple and one
orange,  then again one apple and one orange, etc. This may give
an idea that the number of apples is equal to the number of
oranges. But one can equally well take each time one apple and
two oranges, and conclude that the number of oranges is twice as
large as the number of apples. The main problem here is that we
are making an attempt to compare two infinities, and this gives
an ambiguous result. Similarly, the total volume of a
self-reproducing inflationary universe diverges in the future.
When we make slices of the universe by hypersurfaces of constant
time $t$, we are choosing one particular way of sorting out this
infinite volume. If one makes the slicing in a different way,
the results will be different. This forces us to be very
cautious when using various probability distributions in quantum
cosmology \cite{GBLL}.

Nevertheless, it is very tempting to consider various
cosmological problems using the probability distribution $P_p$
as a guide. In many cases one can get simple and unambiguous
results. In some other cases, especially when one investigates
speculative possibilities related to the baby universe theory
and the choice between different coupling constants, one should
not take the corresponding results too seriously. This being
said, it would be most interesting to see whether this approach
is capable, at least in principle,  to give us any new insights
into such profound problems as the cosmological constant
problem.

The first attempts to study this question were not very
enlightening. The best result which one could obtain was to
reduce the  interval of possible values of the vacuum energy
$V_0$ from  $-10^{94} g \cdot cm^{-3} \leq V_0  \leq 10^{94} g
\cdot cm^{-3}$ to  $-10^{-29} g \cdot cm^{-3} \leq V_0  \leq
10^{-27} g \cdot cm^{-3}$ \cite{CosmConst}. This was achieved by
justifying anthropic bounds in the context of inflationary
cosmology.  Note that the constraint $-10^{-29} g \cdot cm^{-3}
\leq V_0$ follows from the fact that the universe with the
negative vacuum energy $V_0  < -10^{-29} g \cdot cm^{-3}$ would
collapse within $10^{10}$ yrs. This constraint does not differ
very much from the observational constraints on the vacuum
energy $|V_0| \leq 10^{-29} g \cdot cm^{-3}$. The  anthropic
constraint  $V_0  \leq 10^{-27} g \cdot cm^{-3}$ on the {\it
positive} cosmological constant follows from the theory of
galaxy formation  \cite{Weinberg}. Unfortunately, it allows the
vacuum energy to be about two orders of magnitude greater than
$10^{-29} g \cdot cm^{-3}$. This disagreement remains the most
difficult part of the cosmological constant problem. We will
argue in the present paper, that under certain conditions this
part of the problem can be resolved.

Another problem to be discussed is related to the recent
argument of ref.  \cite{Vil,Al} that in the context of
inflationary quantum cosmology it is most probable that the
density perturbations are produced by topological defects.  This
argument is  related to the baby universe theory, the
possibility to choose between the theories with different
coupling constants, and the probability distribution $P_p$.  We
will try to formulate this argument in a more exact form and
examine its validity, taking into account the results of ref.
\cite{LLM}.

In order to compare theories with different coupling constants
(and different values of the cosmological constant) in the
context of quantum cosmology, one should know first of all how
inflation can be realized in each of them. In particular,
self-reproduction of the universe and stationarity of the
probability distribution $P_p$ are not generic properties of all
inflationary models. There are some inflationary models where
self-reproduction does not occur, while there are other in which
it does, but the probability distribution $P_p$ is not
stationary: it constitutes what we called a runaway solution
\cite{GBLL,JGB}. In the main part of the present paper we will
consider a large class of inflationary models where each of
these regimes can be realized.

In Section \ref{minimal} we will discuss the main features of
the chaotic inflation scenario in the simple theory of a scalar
field $\phi$ minimally coupled to gravity, with the effective
potential ${\lambda\over 4}\phi^4$ \cite{b17}. In this
discussion we will follow refs. \cite{b19,LLM}.  In Section 3 we
will consider the same model, but with the scalar field  $\phi$
nonminimally coupled to gravity due to the term ${1\over
2}\xi\phi^2 R$ in the Lagrangian. Models of this type have been
extensively studied by many authors, see e.g.
\cite{Spokoiny,Maeda}.  However, the theory of self-reproduction
of the universe and the behavior of the distribution $P_p$ in
these models have not been addressed so far. Meanwhile, as we
will see, this behavior can be quite nontrivial. Depending on
the value of the  coupling constant $\xi$, each of the regimes
mentioned in the previous paragraph can be realized in these
models. In Section 4 we will generalize this model by including
one-loop quantum gravity corrections (conformal anomaly). The
model we will consider is a hybrid of the standard chaotic
inflation scenario (with an arbitrary coupling ${1\over
2}\xi\phi^2 R$) and the Starobinsky model \cite{Star1}. We will
show, in particular, that one of the inflationary branches in
Starobinsky model, which had been considered unphysical for the
reason that the Hubble constant on this branch was growing
rather than decreasing,  may have a very interesting
interpretation when taking into account the self-reproduction
of the inflationary universe.

The results obtained in Sections 2--4 are of some interest
independently of the speculative discussion contained in
the last  part of our paper, Section \ref{CosmConst}, where we
compare various theories in the context of quantum cosmology.
To avoid possible misunderstandings, we should emphasize from
the very beginning that  in the present paper this discussion is
carried out in the context of   the baby universe theory. We
will compare different quantum states of the universe (we will
call them different ``universes''), which are described by
theories with different coupling constants.  This approach
differs considerably from the more conventional approach
developed in \cite{LLM,GBLL},  where  different exponentially
large causally disconnected parts of the same universe, which
may have different laws of low-energy physics inside each of
them, have been compared to each other. The reason for this
difference is that the total volume of different universes may
grow at a different rate, depending on the coupling constants,
vacuum energy, etc. On the other hand, as  it was shown in
\cite{LLM,GBLL}, the total volume of all parts of the universe
described by a stationary probability distribution $P_p$ grows
at the same
rate. Therefore when comparing different universes the main
effect may arise from comparing the rates of expansion for
various values of coupling constants. Meanwhile, when
considering different parts of the same universe, the relative
fraction of the volume of a stationary self-reproducing universe
in a given state (i.e. in a state with given fields, given effective
coupling constants, etc.) is controlled by the normalized
probability distribution $P_p$.  Comparing different parts of
the same universe has a much simpler interpretation that the
speculative possibility of comparing different quantum
states of the universe. However, we believe that some
``theoretical experiments'' with the baby universe theory may be
useful, since they allow us to look at many problems of quantum
cosmology from a new perspective. In particular, in Section
\ref{CosmConst} we will consider inflation in  the  Starobinsky
model. We will    argue that if the cosmological constant in
this model is non-negative, then it is most probably zero.

\section{\label{minimal} Elementary chaotic inflation model}

In this section we will describe the classical evolution of the
inflaton field with a generic chaotic potential, $V(\phi) =
{\lambda\over4}\,\phi^4$,
\begin{equation}\label{SFF}
{\cal S}=\int d^4x \sqrt{-g} \left[{M_{\rm p}^2\over16\pi} R
- {1\over2}(\partial\phi)^2 - V(\phi)\right]\ .
\end{equation}
The equations of motion during inflation  can be written as
follows:
\begin{equation}\begin{array}{rl}\label{EMS}
&{\displaystyle H = \left({8\pi V(\phi)\over3M_{\rm p}^2}
\right)^{1/2} = \left({2\pi\lambda\over3}\right)^{1/2}
{\phi^2\over M_{\rm p}} \ ,}\\[4mm]
&{\displaystyle \dot\phi = - {V'(\phi)\over3H} = -
\left({\lambda\over6\pi}\right)^{1/2} M_{\rm p}\,\phi }\ .
\end{array}\end{equation}
According to these equations, the inflationary regime ($|\dot H|
< H^2$)
occurs at $\phi > \phi_{\rm e}$, where $\phi_{\rm e}
= M_{\rm p}/\sqrt\pi$. However, these equations are valid
only at densities smaller than the Planck density, $V(\phi) <
M_{\rm p}^4$, or $\phi < \phi_{\rm p} = (\lambda/4)^{-1/4}\,
M_{\rm p}$.  We will call $\phi_{\rm e}$ the end of inflation
boundary, and $\phi_{\rm p}$ the Planck boundary.

The inflaton field fluctuates in de Sitter space during a time
interval $\Delta t = H^{-1}$ with amplitude approximately equal
to the Gibbons--Hawking temperature,
\begin{equation}\label{STEP}
\delta\phi = {H\over2\pi}\ .
\end{equation}
Quantum fluctuations then act on the coarse-grained background
field as stochastic forces, producing a Brownian motion on the
value of the inflaton field. In most of the inflationary
domains, the inflaton field will follow the classical evolution
towards the end of inflation. However, those rare domains in
which the inflaton field grows due to quantum fluctuations will
inflate more, since the rate of expansion $H$ is proportional to
$\phi^2$.  Beyond a certain value of the inflaton field $\phi =
\phi_{\rm s}$, for which the amplitude of quantum fluctuations
becomes larger than its change due to classical motion in the
same time interval, $\delta\phi > \Delta\phi = \dot\phi H^{-1}$,
we enter the regime of self-reproduction of the universe. In the
case of a simple quartic potential, such a value is given by
$\phi_{\rm s} = (2\pi \lambda/3)^{-1/6}\,M_{\rm p}$.

The Brownian motion of the inflaton during the self-reproduction
of the universe can be described in the physical frame, which
takes into account the growth of the proper volume of the
inflationary domain, with an ordinary diffusion equation
\cite{Nambu,LLM},
\begin{equation}\label{FPP}
{\partial{\cal V}\over\partial t} = {\partial\over\partial\phi}
 \left({H^{3/2}\over 8\pi^2} {\partial\over\partial\phi}
\left({H^{3/2}}{\cal V}\right) + {V'\over3H}\,{\cal V}\right)
+ 3H {\cal V} \  .
\end{equation}
where ${\cal V}(\phi,t)$ is the total volume of all domains
containing scalar field $\phi$. In the terminology of ref.
\cite{LLM}, this is the non-normalized probability distribution
$P_p$.

It is possible to find solutions to this equation subject to the
appropriate boundary conditions at the Planck boundary and the
end of inflation. Such solutions are generically of the form
\begin{equation}\label{SOL}
{\cal V}(\phi,t) = e^{\alpha t}\ P_p(\phi)\ .
\end{equation}
where $\alpha$ is some constant, and $P_p$ is a time-independent
normalized probability distribution to find a field $\phi$ in a
unit physical volume. In case that the dependence of ${\cal
V}(\phi,t)$ on $\phi$ and $t$ can be factorized as in eq.
(\ref{SOL}), we will speak about stationary solutions for ${\cal
V}(\phi,t)$ and for $P_p(\phi)$. Dependence of these solutions
on the conditions at the boundary where inflation ends is
exponentially weak \cite{LLM}. However, in general, solutions
strongly depend on the boundary conditions at large $\phi$.  The
simplest boundary condition one can impose is ${\cal V}
(\phi_{\rm p}) = 0$.  One can argue that inflation ceases to
exist at $V(\phi) > M^4_{\rm p}$ because of large quantum
fluctuations \cite{LLM}. An advanced version of this argument
was recently given in \cite{Vil}; we will consider it in Section
4. One can show that the stationary solution for ${\cal V}$ with
the boundary condition ${\cal V} (\phi_{\max}) = 0$ (whatever is
the value of the $\phi_{\max} \gg \phi_{\rm e}$) is given by
\cite{LLM}
\begin{equation}\label{SOL2}
{\cal V}(\phi,t) \sim e^{d(\lambda) H_{\max}(\lambda) t}\
P_p(\phi)\ .
\end{equation}
The coefficient $d(\lambda)$  in the chaotic inflation scenario
can be interpreted as a fractal dimension of inflationary
domains  at the upper boundary $\phi = \phi_{\max}$ \cite{LLM}.
(For a   discussion of the fractal dimension in the context of
the new inflationary scenario see \cite{ArVil}.)

If the upper  boundary  $\phi_{\max}$ coincides with the Planck
boundary $\phi_{\rm p}$  defined by the condition $V(\phi_{\rm
p}) = M_{\rm p}^4$, then the distribution ${\cal V}$ grows as
follows:
\begin{equation}\label{SOL2a}
{\cal V}(\phi,t) \sim e^{(3-f(\lambda))\sqrt{8\pi\over 3}
M_{\rm p} t}\ P_p(\phi)\ .
\end{equation}
where
$f(\lambda) = 3 -d(\lambda)$. Another useful form of eq.
(\ref{SOL2a}) is
\begin{equation}\label{SOL2b}
{\cal V}(\rho,t) \sim e^{(3-f(\lambda))\sqrt{8\pi\over 3}
M_{\rm p} t}\ P_p(\rho)\ ,
\end{equation}
where ${\cal V}(\rho,t)$ is the total volume of all domains with
a given density and $P_p(\rho)$ is the probability distribution
to find a domain of a unit volume containing matter with density
$\rho$.  If one divides these equations by the
$\phi$-independent factor $e^{(3-f(\lambda))\sqrt{8\pi\over3}
M_{\rm p} t}$, one obtains the normalized stationary probability
distribution $P_p$ discussed in \cite{LLM}. In the present
paper, however, we will often talk about the unnormalized
distributions (\ref{SOL2}), since they show the growth of the
total volume of all domains filled with a given field $\phi$.

It is important to study how the fractal dimension $d(\lambda)$
depends on the coupling constant $\lambda$ \cite{LLM}:
\vspace{-0.5cm}
\begin{center}
\begin{tabular}{|c|c|c|c|c|c|c|c|}
\hline \hline
$\lambda$ & $1$ & $10^{-1}$ & $10^{-2}$ & $10^{-3}$ & $10^{-4}$
& $10^{-5}$ & $10^{-6}$ \\
\hline
 $d$ & 0.9719 & 1.526 & 1.915 & 2.213 & 2.438 & 2.604 & 2.724
\\
\hline \hline
\end{tabular}
\end{center}
\vspace{-0.1cm}
As we see, $d(\lambda)$ grows with decreasing $\lambda$ towards
the usual space dimension $3$.
This means that $f(\lambda)$ decreases with $\lambda$; it can be
shown that $f(\lambda)$  vanishes in the limit $\lambda \to 0$.

Note  that the distributions ${\cal V}(\phi,t)$ and
$P_p$  depend on the choice of time parametrization. For
example,  instead of  usual time $t$ measured by observers by
their clock, one can use ``time'' $\tau  = \ln{{a\left(x, t
\right) \over a(x,0)}} =  \int{H(\phi(x,t),t)\ dt}$.  Here
$a\left(x, t \right)$ is a local value of the scale factor in
the inflationary Universe. Obviously, the time $\tau$ measures
the logarithm of the local expansion of the universe. Solution
of the diffusion equation for ${\cal V}(\phi,\tau)$ is also
stationary, but it looks slightly different \cite{LLM},
\begin{equation}\label{SOL2c}
{\cal V}(\phi,\tau) \sim e^{(3-1.1\sqrt\lambda)\tau}\
\tilde P_p(\phi)\ ,
\end{equation}
which corresponds to a fractal dimension $3-1.1\sqrt\lambda$.
It what follows it will be important for us, that even though
the distributions ${\cal V}(\phi,t)$ and ${\cal V}(\phi,\tau)$
differ from each other, both share the same property: they grow
at large $t$ (at large $\tau$) with a rate which increases as
$\lambda$ goes to zero.

One should emphasize  \cite{LLM}  that the factor $e^{\alpha t}
\sim e^{d(\lambda) H_{\max}(\lambda)  t}$ in (\ref{SOL2})  (as
well as the factor $e^{(3-1.1\sqrt\lambda)\tau}$ in
(\ref{SOL2c})) gives the rate of growth of the combined volume
of all domains with a given field $\phi$ (or of all domains
containing matter with a given density) {\it  not only at very
large $\phi$, where quantum fluctuations are large, but at small
$\phi$ as well, and even after inflation} \cite{diff}.This
result may seem absolutely unexpected, since the volume of each
particular inflationary domain grows like $e^{3H(\phi)t}$, and
after inflation the law of expansion becomes completely
different.  One should distinguish, however, between the growth
of each particular domain, accompanied by a decrease of density
inside it, and the growth of the total volume of all domains
containing matter with a given (constant) density. In the
standard big bang theory the second possibility did not exist,
since the energy density was assumed to be  the same in all
parts of the universe (``cosmological principle''), and it was
not constant in time.

The reason why there is a universal expansion rate (\ref{SOL2})
can be understood as follows. Because of the self-reproduction
of the universe there always exist many domains with $\phi \sim
\phi_{\max}$,  and their combined volume grows almost as fast as
$e^{3H_{\max} t}$. Then the field $\phi$ inside some of these
domains decreases. The  total volume of domains containing some
small field $\phi$ grows not only due to expansion $\sim
e^{3H_{\max} t}$, but mainly due to the unceasing process of
expansion of domains with large $\phi$ and their subsequent
rolling (or diffusion) towards small $\phi$.  The above
mentioned universality of the expansion law will play a crucial
role in our discussion of quantum cosmology in the Sections
\ref{CosmConst}, \ref{Discussion}. In what follows we will turn
to  other models, where this law may or may not hold.

\section{\label{model2} Nonminimal coupling to gravity}

In this section we will describe the classical evolution of the
inflaton field with a generic chaotic potential, coupled to the
curvature scalar with a small non minimal coupling $\xi$,
\begin{equation}\label{S2}
{\cal S}=\int d^4x \sqrt{-g} \left[{M_{\rm p}^2\over16\pi} R
- {1\over2} \xi \phi^2 R - {1\over2}(\partial\phi)^2 -
V(\phi)\right]\ .
\end{equation}
In this theory the Planck mass takes the form
\begin{equation}\label{PM1}
M_{\rm p}^2(\phi) = M_{\rm p}^2 - 8\pi \xi\phi^2 \ .
\end{equation}
We can write the equations of motion for the homogeneous field
$\phi$ during inflation, in the slow roll-over approximation, as
\begin{equation}\begin{array}{rl}\label{SEM}
{\displaystyle R\ \simeq 12H^2\ }=
&{\displaystyle {32\pi V(\phi)\over M_{\rm p}^2-8\pi\xi\phi^2}\
,}\\[4mm]
(1 - 6\xi)\ 3H\dot\phi\phi\ = &{\displaystyle 4V(\phi)-\phi
V'(\phi) - {M_{\rm p}^2\over8\pi}\ R}\ .
\end{array}\end{equation}
We will consider two different cases, $\xi>0$ and $\xi<0$, with
$|\xi| \ll 1/6$.

{\bf 1.\ Case $\xi>0$}.
In this case there is a clear bound on the inflaton field,
\begin{equation}\label{CRI}
\phi < \phi_{\rm c} \equiv {M_{\rm p}\over\sqrt{8\pi\xi}}\ ,
\end{equation}
in order that gravity be attractive, see eq. (\ref{PM1}). We
will consider a typical potential of chaotic inflation,
$V(\phi)=\lambda\phi^4/4$.  The equations of motion (\ref{SEM})
are then
\begin{equation}\begin{array}{rl}\label{SEM2}
3H^2\ = &{\displaystyle {2\pi\lambda\phi^4\over M_{\rm p}^2 -
8\pi\xi\phi^2} \ ,}\\[4mm]
(1 - 6\xi)\ 3H\dot\phi\phi\ = &{\displaystyle -\ {\lambda\phi^4
M_{\rm p}^2\over M_{\rm p}^2 - 8\pi\xi\phi^2}\ .}
\end{array}\end{equation}
The Planck boundary is given by
\begin{equation}\label{PB2}
\phi_{\rm p}^2 = {2M_{\rm p}^2\over\sqrt\lambda+16\pi\xi} \ .
\end{equation}
Note that $\phi_{\rm p} < \phi_{\rm c}$; in particular,
$\phi_{\rm p} = \phi_{\rm c}\Bigl(1 - {\sqrt\lambda \over
32\pi\xi}\Bigr)$ for $\sqrt\lambda \ll 16\pi\xi$.

Therefore, if $\sqrt\lambda < 16\pi\xi \ll 1$, the inflaton
field will stop just before $\phi_{\rm c}$. On the other hand,
the range of inflation corresponds to $|\dot H| < H^2$,
which gives
\begin{equation}\label{END}
{M_{\rm p}^2\over\pi} < \phi^2 < \phi_{\rm i}^2 \ ,
\end{equation}
where the upper boundary of the inflationary regime is given by
$\phi_{\rm i} = \phi_{\rm c}\,\left(1-2\xi\right)$. Therefore,
in order that inflation ends before reaching Planck boundary
(\ref{PB2}), we require $\sqrt\lambda < 64\pi\xi^2$.  The
classical motion of the coarse-grained inflaton field is
affected by its quantum fluctuations, with amplitude
\begin{equation}\label{AMP}
\delta\phi = {H\over2\pi}(1-6\xi)^{-1/2}\ ,
\end{equation}
where the extra factor comes from the proper normalization of
the inflaton field.

An inflationary domain of the universe will enter the
self-reproduction regime when the amplitude of quantum
fluctuations of the inflaton field in that domain $ \delta\phi$
is larger than the corresponding classical motion $\Delta\phi =
\dot\phi H^{-1}$ in the interval $\Delta t = H^{-1}$. This
happens for $\phi > \phi_{\rm s}$, where
\begin{equation}\label{SELF2}
\phi_{\rm s}^2 =
\phi_{\rm c}^2\,\left(1 - {\lambda\over 768\pi^2\xi^3}\right) \ .
\end{equation}
Domains with $\phi > \phi_{\rm s}$ will inflate and produce more
domains with even higher values of the inflaton, until they
reach the upper boundary of inflation $\phi_{\rm i}$.

To investigate various regimes which are possible in the model
(\ref{S2}) with $\xi > 0$, one should compare
$\phi_{\rm e}$, $\phi_{\rm i}$, $\phi_{\rm s}$ and $\phi_{\rm
p}$ for various relations between $\lambda$ and $\xi$. This
shows that for $\xi > 1/6$ there is no inflation, and for $\xi
\ll 1/6$ there exist three different regimes:

\begin{itemize}
\item [1.] $0 < \xi < {\displaystyle {\lambda^{1/2}\over16\pi}}$.
We recover the usual results of chaotic inflation with a quartic
potential.
\item [2.] ${\displaystyle {\lambda^{1/2}\over16\pi} < \xi <
{\lambda^{1/4}\over8\sqrt\pi}}$. There is inflation and
self-reproduction.
\item [3.] ${\displaystyle {\lambda^{1/4}\over8\sqrt\pi} < \xi
< {1\over6}}$. There is inflation but no self-reproduction.
\end{itemize}

In this case there are no runaway solutions since the quantum
diffusion of the inflaton stops just before the critical value
(\ref{CRI}).  For arbitrary potentials we may or may not have
inflation, but there will never be runaway solutions, even if
the parameters allow for a Planck boundary, because of the bound
(\ref{CRI}).

{\bf 2.\ Case $\xi < 0$.}
The Planck mass in this case is given by
\begin{equation}\label{PM2}
M_{\rm p}^2(\phi) = M_{\rm p}^2 + 8\pi|\xi|\phi^2 \ .
\end{equation}
Thus, it is always positive, and there is no bound on $\phi$.
In fact, for large values of the inflaton field, $M_{\rm
p}(\phi)$ will become dominated by its evolution.

Let us consider again $V(\phi)=\lambda\phi^4/4$, and later
comment on alternatives. The equations of motion are
\begin{equation}\begin{array}{rl}\label{SEM4}
3H^2\ = &{\displaystyle {2\pi\lambda\phi^4\over M_{\rm p}^2 +
8\pi|\xi|\phi^2} \ ,}\\[4mm]
(1 + 6|\xi|)\ 3H\dot\phi\phi\ = &{\displaystyle -\ {\lambda\phi^4
M_{\rm p}^2\over M_{\rm p}^2 + 8\pi|\xi|\phi^2}\ .}
\end{array}\end{equation}
The Planck boundary is given by
\begin{equation}\label{PB4}
\phi^2 (\sqrt\lambda - 16\pi|\xi|) < 2M_{\rm p}^2\ .
\end{equation}
Therefore, if $\sqrt\lambda < 16\pi|\xi|$, then the inflaton
will never reach Planck density. On the other hand, the end of
inflation for small $|\xi|$, is given by $\phi_{\rm e} = M_{\rm
p}/\sqrt\pi$.

The amplitude of quantum fluctuations of the normalized inflaton
field is given by (\ref{AMP}) with $\xi = - |\xi|$.
Self-reproduction in this case will occur, for small $|\xi|$, at
\begin{equation}\label{SELF4}
{\delta\phi\over\Delta\phi} = {H\phi\over M_{\rm p}^2}
(1 + 6|\xi|)^{1/2} \simeq \left({2\pi\lambda\over3}\right)^{1/2}
{\phi^3\over M_{\rm p}^3} > 1\ .
\end{equation}

There are runaway solutions in this case, since the quantum
diffusion of the inflaton has no boundary for large $\phi$, see
eq. (\ref{PB4}), and the probability distribution will move
towards the maximum expansion rate. For large $\phi$, this
becomes
\begin{equation}\label{PRO}
H \simeq \sqrt{\lambda\over 12|\xi|} \ \phi \ ,
\end{equation}
while the value of Planck mass also grows with $\phi$, as
$M_{\rm p} = (8\pi|\xi|)^{1/2}\,\phi$.

Runaway solutions are probability distributions satisfying a
diffusion equation similar to (\ref{FPP}) that move forever
towards large values of the field $\phi$ cite{GBLL,JGB}. The way it moves will
depend on the type of potential.  For $\lambda\phi^4$ it is an
explosive behavior, as we will see. The probability distribution
gives a statistical description of the quantum diffusion process
towards large $\phi$, but it proves useful to analyze the
particular behavior of those relatively rare domains in which
the field $\phi$ increases in every quantum jump of amplitude
(\ref{STEP}).  We can compute the speed at which those domains
move towards large values of the field $\phi$ from
\begin{equation}\label{STH}
\dot\phi = {\delta\phi\over\Delta t} = {H^2\over2\pi}
= {\lambda\over24\pi|\xi|}\,\phi^2\ ,
\end{equation}
where $H$ is given by (\ref{PRO}). We find an explosive solution
\begin{equation}\label{SST}
\phi(t)=\phi_0\ \left(1 - {\lambda\over24\pi|\xi|}\,\phi_0\,t
\right)^{-1}\ .
\end{equation}
We see that those first domains of the diffusion process reach
infinity in finite time. Note that the total volume of such
domains at that time will be finite, and then they will start
growing at an infinitely large rate.  This behavior is explosive,
it corresponds to probability distributions that are
nonstationary and singular at $\phi \to \infty$.

It should be noted that the existence of runaway solutions or
even inflation is not generic. In fact, for arbitrary potentials
of the type $V(\phi) = {\lambda\over2n} \phi^{2n}$, $n>2$, the
equations of motion (\ref{SEM}) become, for large $\phi$,
$\,8\pi|\xi|\phi^2 \gg M_{\rm p}^2$,
\begin{equation}\begin{array}{rl}\label{EM3}
\dot\phi\ = &{\displaystyle -\,\left({6|\xi|\over1+6|\xi|}\right)
{H\phi\over3}\ (n-2)\ ,}\\[3mm]
H\ = &{\displaystyle \left({\lambda\over6n|\xi|}\right)^{1/2}
\phi^{n-1}}\ .
\end{array}\end{equation}
It can easily be seen that there is inflation ($|\dot H| < H^2$)
only for
\begin{equation}\label{NN1}
(n-1)(n-2) < 3\,\left({1+6|\xi|\over6|\xi|}\right)\ .
\end{equation}
In particular, for an exponential potential $n=\infty$, the non
minimally coupled term prevents inflation at large $\phi$, since
it produces a very rapid motion of the scalar field.
Furthermore, one can also compute the self-reproduction regime
for an arbitrary chaotic potential like above. For those values
of $n$ for which there is inflation, the condition $\dot\phi
H^{-1} < \delta\phi$ reads,
\begin{equation}\label{SELF}
4\pi\sqrt n\,(n-2)\,{|\xi|\over\sqrt\lambda}\,
\left({6|\xi|\over1+6|\xi|}\right)^{1/2} <
\left({\phi\over M_{\rm p}}\right)^{n-2} <
8\pi\sqrt{2n}\,{|\xi|\over\sqrt\lambda}\ ,
\end{equation}
where the last inequality comes from the Planck boundary in the
large inflaton limit. It is clear that in order to have
self-reproduction we need
\begin{equation}\label{NN2}
(n-2)^2 < 8\,\left({1+6|\xi|\over6|\xi|}\right)\ .
\end{equation}
In conclusion, as we increase $|\xi|$, first self-reproduction
disappears, and then even inflation itself is no longer
sustained for a given $n$. Therefore, runaway solutions are very
special and appear only for theories satisfying both (\ref{NN1})
and (\ref{NN2}).

\section{\label{model3} Chaotic inflation and conformal anomaly}
\subsection{Starobinsky model with inflaton field}
In this section we will describe inflation in the combined
model, including scalar fields and the $R^2$ terms, which may
appear in the theory because of the one-loop quantum gravity
effects.  Note however, that these terms by themselves can lead
to the existence of inflationary regime, as was first realized
by Starobinsky \cite{Star1}.  Therefore in this subsection we
will remember some basic features of the Starobinsky model.
After that we will add a non minimal coupling to the inflaton
field.

The equations of motion associated with Starobinsky model in the
presence of an inflaton field with potential $V(\phi)$ can be
written as follows:
\begin{equation}\begin{array}{rl}\label{XE3}
\nabla^2\phi\ = &{\displaystyle -\ V'(\phi) \ ,}\\[3mm]
{\displaystyle -{M_{\rm p}^2\over8\pi}\left(R_{\mu\nu} -
{1\over2}g_{\mu\nu} R\right)\ =}
&{\displaystyle \ g_{\mu\nu} V(\phi)\ +\ \partial_\mu\phi
\partial_\nu\phi - {1\over2} g_{\mu\nu} (\partial\phi)^2 }\\[3mm]
&{\displaystyle +\ {M_{\rm p}^2\over8\pi}
\left({1\over6M^2}\ ^{(1)}H_{\mu\nu} +
{1\over H_0^2}\ ^{(3)}H_{\mu\nu}\right)}\ ,
\end{array}\end{equation}
where
\begin{equation}\begin{array}{rl}\label{HIJ}
^{(1)}H_{\mu\nu}\ = &{\displaystyle 2 \left(\nabla_\mu\nabla_\nu
- g_{\mu\nu} \nabla^2\right) R + 2 R R_{\mu\nu} -
{1\over2} g_{\mu\nu} R^2\ ,}\\[3mm]
^{(3)}H_{\mu\nu}\ = &{\displaystyle R_\mu^{\ \lambda}
R_{\lambda\nu} - {2\over3} R R_{\mu\nu} - {1\over2} g_{\mu\nu}
R^{\rho\sigma} R_{\rho\sigma} + {1\over4} g_{\mu\nu} R^2}\ .
\end{array}\end{equation}
The parameters $H_0$ and $M$ in the original version of the
Starobinsky model were related to the conformal anomaly, but in
a later version the term ${R^2\over6M^2}$ was simply added to
the Lagrangian by hand \cite{Star2,KLS}. The value of $H_0$ is
of the same order as $M_{\rm p}$, but it can be somewhat smaller
if there are many matter field (of spin 0, 1/2 and 1)
contributing to the conformal anomaly. In fact, one loop
gravitational corrections in our theory are somewhat more
complicated, especially because the theory of the inflaton field
$\phi$ is not conformally invariant. Nevertheless when the
number of other fields contributing to conformal anomaly is
sufficiently large, i.e. when the masses $M^2$ and $H_0^2$ are
sufficiently small, our approximation may be reasonable.

During inflation one can write the equations of motion for the
homogeneous fields $\phi$ and $H$, in the slow rolling
approximation, as
\begin{equation}\label{RS3a}
\dot H = - {M^2\over6}\,\left(1 - {H^2\over H_0^2}
- {8\pi V(\phi)\over 3M_{\rm p}^2 H^2}\right)\ ,
\end{equation}
\begin{equation}\label{RS3b}
\dot\phi\ =  - {V'(\phi)\over3H}  \ .
 \end{equation}
Let us neglect the scalar field first, i.e. consider the
original Starobinsky model first.  All possible inflationary and
non-inflationary regimes in this model have been described in  a
particularly detailed way in \cite{KMP}.  However, as we will
see, with an account taken of the self-reproduction of
inflationary universe  we can get some additional (and rather
unexpected) information about this model.

First of all, let us remember that there exist three different
inflationary regimes in this model, and only two of them are
usually considered in the literature.  Namely, the first stage
of inflation occurs  for $H^2_0 \gg M^2$ in the regime with
$|\dot H| < M^2/6 \ll H_0^2$.
In the limiting case $\dot H = 0$ inflation occurs with the
Hubble constant $H = H_0$.  Just as in the new inflation model,
this regime is unstable, and inflation enters the second regime
with $\dot H$ asymptotically approaching $- M^2/6$.
During this process the Hubble constant decreases, until it
reaches the value $H \sim M$, which corresponds to the end of
inflation.

However, there also exists another, rather unusual inflationary
branch. Namely, eq. (\ref{RS3a}) at $V(\phi) = 0$, $H \gg H_0$,
reads
\begin{equation}\label{RS3c}
\dot H =  {M^2\over6H_0^2}\,H^2\ .
\end{equation}
Obviously, this is an inflationary regime with $|\dot H| \ll
H^2$ for $M^2 \ll 6H_0^2$.
In this regime the Hubble
constant indefinitely grows, and approaches infinitely large
values within a finite time $\Delta t \sim 6H_0/M^2$. For this
and some other reasons this branch was believed to be unphysical
and not very interesting \cite{KMP,Vil}.
Neglecting the possibility of
inflation with positive $\dot H$ can lead to important
constraints on the rate of inflation in the Starobinsky model.

Indeed, equation (\ref{RS3a}) in the limit
$\dot H \to 0$ has two possible solutions \cite{Chibisov}:
\begin{equation} \label{RS3}
H^2\ =   {H_0^2\over2} \left(1 \pm \sqrt{1 - {32\pi V(\phi)\over
3H_0^2 M_{\rm p}^2}}\,\right) \ .
\end{equation}
In the limit of small $V(\phi)$, these two branches obviously
correspond to the Starobinsky inflation with $H^2 = H_0^2$, and
to the scalar field driven inflation with $H^2 = {8\pi
V(\phi)\over 3 M_{\rm p}^2}$. One can easily see, that for $\dot
H < 0$ the Hubble constant $H$ on the upper branch becomes
smaller than $H_0$.
Thus, one can consider $H_0$ as an upper bound on the rate of
inflation in a rather general class of models. Typically this
bound is very close to the Planck bound
$H_{\rm max} \sim M_{\rm p}$, but in certain cases $H_0$ can be
somewhat smaller than $M_{\rm p}$  \cite{Vil}. This is a very
interesting observation, since it provides a natural upper
boundary which is necessary for finding the probability
distribution $P_p$ \cite{LLM}.  For $H_0 \ll M$ the upper
boundary for inflation becomes even lower.  In this case there
is no inflation close to the   upper branch of eq. (\ref{RS3}),
and the lower branch   cannot go higher than $H_{\rm max}  \sim
H _0/\sqrt 2$, which corresponds to $V(\phi)
\sim {3H_0^2 M_{\rm p}^2\over32\pi}$.

However, if inflation near the upper branch of  (\ref{RS3}) is
possible, then one cannot always neglect the possibility of
inflation with $\dot H > 0$.  Indeed, in the usual chaotic
inflation models classical motion of the scalar field shifts it
to smaller values of $V(\phi)$. However, quantum jumps of the
field $\phi$ in the regime of self-reproduction can move it
against the classical flow, towards the highest possible values
of $V(\phi)$. Similarly, the classical motion shifts $H$ towards
the singularity at the inflationary trajectory with $\dot H > 0$.
However, if this trajectory allows self-reproduction, the Hubble
constant $H$ may drift towards its very large values, and then
in some of inflationary domains it may jump back towards $H <
H_0$. Since the rate of expansion of the universe at the branch
with $H > H_0$ is very large, the volume of the corresponding
parts of the universe grows at a very high rate, and the
existence of the ``pathological'' inflationary branch with $H >
H_0$ will give a dominant contribution to the overall rate of
the universe expansion $e^{dH_{\rm max} t}$. In this case
$H_{\rm max}$ will   be greater than $H_0$. One can still argue
that $H_{\rm max}$ should not be much greater than $M_{\rm p}$,
since in this case the notion of classical space-time ceases to
exist, and the usual derivation of the stochastic equations for
$P_p$ becomes invalid \cite{LLM}.

In order to study the process of the universe self-reproduction
in  the Starobinsky model, we should find the  amplitude of
quantum fluctuations of the scalar curvature $R \sim 12 H^2$,
which are related in the following way to the fluctuations of
the canonically normalized scalaron field $\delta\varphi$:
\begin{equation}\label{QFL}
\delta R = (48\pi C)^{1/2} \,{M^2\over M_{\rm p}}\,
\delta\varphi\ ,
\end{equation}
where $C = 1 + R/3M^2 - R/6H_0^2$ \cite{Chibisov}.  Both the
scalaron $\delta\varphi$ and the inflaton fluctuations
$\delta\phi$ satisfy approximate massless equations in de Sitter
space, whose solution is well known, and whose amplitude is
approximately given by Gibbons-Hawking temperature, $H/2\pi$.
(Scalaron has a tachyonic mass squared $-M^2$, where $M^2 \ll
H^2$ \cite{Chibisov}.) Using (\ref{RS3}) we find $C^{1/2}\simeq
2H/M$, and therefore
\begin{equation}\begin{array}{rl}\label{QF}
\delta\varphi\ =& {\displaystyle \delta\phi\  =  {H\over2\pi}
\ ,}\\[3mm]
\delta H\  =&{\displaystyle {H\over2\pi}\ \sqrt{\pi\over3}\
{M\over M_{\rm p}}}\ .
\end{array}\end{equation}

Let us now consider inflation at $H \simeq H_0$, with $V(\phi)
\simeq 0$.  Self-reproduction will occur in this case if the
classical shift of the Hubble constant within the time $H^{-1}$
is smaller than the amplitude of fluctuations of $H$. This
condition implies  $|{\dot H\over H}| <\delta H$. Together with
eqs. (\ref{RS3a}, \ref{QF}), this condition gives
\begin{equation}\label{SEF}
\left|1 - {H_0^2\over H^2}\right| < \sqrt{3\over\pi}\,{H_0^2\over
M M_{\rm p}}\ .
\end{equation}
Some care should be taken when applying this criterion at $H
\simeq H_0$: it should be satisfied for all $H$ within the
interval $\delta H$ from $H_0$. This gives the following
criterion of the self-reproduction at $H \simeq H_0$, which
essentially coincides with the criterion for inflation there:
$M \leq \sqrt3 H_0$.
A similar criterion can be obtained from the condition that the
fractal dimension for inflation at $H = H_0$ is positive: $M
\leq 3 H_0$, see eq. (\ref{DIM2}) in the next subsection.

However, this condition is not strong enough to ensure
self-reproduction for $|H- H_0| \geq H_0$. The corresponding
condition follows from (\ref{SEF}).  At small $H$ this condition
reads $M < {H^2\over M_{\rm p}} \sqrt {3\over \pi}$; at large
$H$ this condition is  $M < {H_0^2\over M_{\rm p}} \sqrt {3\over
\pi}$.

Therefore one should consider several different regimes in our
model.

1) $M \geq 3 H_0$.
There is no inflation in the Starobinsky model, but inflation
may exist due to the scalar field $\phi$, for  $V(\phi) <
{3H_0^2 M_{\rm p}^2\over32\pi}$.

2) ${H_0^2\over M_{\rm p}} \sqrt {3\over \pi} < M < 3 H_0$.
There is inflation and self-reproduction at $H \simeq H_0$.
However, self-reproduction occurs only in a narrow band  $H_0
\pm \Delta H$ (\ref{SEF}), with
\begin{equation}\label{IAS}
\Delta H \sim { \sqrt 3\over 2 \sqrt \pi}\,{H_0^3\over
M M_{\rm p}}\ .
\end{equation}
This means, in particular, that only a small part of the branch
with $H > H_0$, $\dot H > 0$ is of any interest for us; the
points which go beyond $H_0 + \Delta H$ never return, and move
towards a singularity. Therefore in this case the effective
maximal value of the Hubble constant will be of the order of
$H_0 + \Delta H$ (\ref{IAS}). After inflation at the upper
branch, the Hubble constant becomes smaller, and inflation
continues at the lower branch, which will be responsible for the
density perturbations in the observable part of the universe.

3) $M < {H_0^2\over M_{\rm p}} \sqrt {3\over \pi}$. In this case
self-reproduction occurs at the whole branch with $H > H_0$,
$\dot H > 0$, at least until inflationary domains  enter the
area where the curvature becomes higher than the Planck one. In
this case there is no upper bound for the Hubble constant near
$H = H_0$; the upper bound for $H$  is expected to be of the
same order as $M_{\rm p}$.

In what follows we will consider the stationary probability
distribution in the case 2), assuming that $H_0 \ll M_{\rm p}$,
${H_0^2\over M_{\rm p}} \sqrt {3\over \pi} \ll M < 3 H_0$. In
this case self-reproduction of the universe occurs in a very
narrow region ($\Delta H \ll H_0$) near $H_0$, see eq. (\ref{IAS}).
The results of our  investigation will be useful for us when we will
discuss the cosmological constant problem.

The corresponding diffusion equation will be written in terms of
the canonically normalized fields $\phi$ and $\varphi$, where
$\varphi$ is the scalaron field introduced in \cite{Chibisov},
see eq. (\ref{QFL}). According to \cite{Chibisov}, this field
satisfies the same equation as a tachyon field with the mass
squared $-M^2$ in de Sitter background. We will assume also that
$V(\phi) = {m^2\over 2 } \phi^2$. Assuming $H \simeq H_0$, one
can show that self-reproduction of the universe occurs in the
intervals
\begin{equation}\begin{array}{rl}\label{SFR}
\phi\ < &{\displaystyle{3H_0^3\over
2\pi\,m^2} \ ,}\\[3mm]
\varphi\ < &{\displaystyle{3H_0^3\over2\pi M^2} \ .}
\end{array}\end{equation}
The diffusion equation for ${\cal V}(\phi,\varphi,t)$ for $H
\simeq H_0$ can be written in the following form:
\begin{equation}\begin{array}{rl}\label{FPE2}
{\displaystyle {\partial {\cal V}\over\partial t}\ }=&
{\displaystyle {H_0^3\over8\pi^2}\,
{\partial^2{\cal V}\over\partial\varphi^2} - {M^2\over3H_0}
{\partial\over\partial\varphi}(\varphi {\cal V}) + 3H_0\left(1 -
{2\pi\phi^2m^2\over3M_{\rm p}^2H_0^2} \right){\cal V} }\\[5mm]
&+ {\displaystyle\ {H_0^3\over8\pi^2}\,
{\partial^2{\cal V}\over\partial\phi^2} +  {m^2\over3H_0}
 {\partial\over\partial\phi}(\phi {\cal V}) }\ .
\end{array}\end{equation}
There is a solution in the limit of large time $t$,
\begin{equation}\label{FPS2}
{\cal V}(\phi,t) \sim \exp(d H_{\max} t)\ \exp\left(-2\pi^2\,a\,
{\phi^2\over H_0^2}\right)\ .
\end{equation}
In our case, the maximum rate of inflation which provides the
largest relative volume is given by $H_{\max} \simeq H_0 +
\Delta H \simeq H_0$. In the limit of quantum diffusion
dominating classical drift, $H_0^2 \gg m M_{\rm p}$, we find
\begin{equation}\label{AMH2}
a \simeq   {m\over\sqrt\pi M_{\rm p}}\ ,
\end{equation}
while the fractal dimension is given by
\begin{equation}\label{DIM2}
d =\ 3 - {M^2\over3H_0^2}  - {m\over 2\sqrt\pi M_{\rm p}}\ .
\end{equation}
The stationary solution (\ref{FPS2}) is a Gaussian centered at
$\phi=0$. It does not depend on $\varphi$ in the domain of
self-reproduction (\ref{SFR}).

\subsection{Starobinsky model with nonminimally coupled scalar
field} In this section we will briefly describe the classical
evolution of the inflaton field with a generic chaotic
potential, coupled to the curvature scalar with a small coupling
$\xi < 0$, $|\xi| \ll 1/6$, in the context of the Starobinsky
model. In this case our equations of motion are somewhat
modified,
\begin{equation}\begin{array}{rl}\label{XEQ}
\nabla^2\phi\ = &{\displaystyle -\ V'(\phi) + |\xi| \phi R\ ,
}\\[3mm]
{\displaystyle - {M_{\rm p}^2(\phi)\over8\pi}
\left(R_{\mu\nu} - {1\over2}g_{\mu\nu} R\right)\ =}
&{\displaystyle \left(\nabla_\mu\nabla_\nu - g_{\mu\nu}
\nabla^2\right) |\xi|\phi^2 \ + \ \partial_\mu\phi
\partial_\nu\phi - {1\over2} g_{\mu\nu} (\partial\phi)^2}\\[3mm]
&{\displaystyle +\ g_{\mu\nu} V(\phi) + {M_{\rm p}^2(\phi)\over
8\pi}\left({1\over6M^2(\phi)}\ ^{(1)}H_{\mu\nu} +
{1\over H_0^2(\phi)}\ ^{(3)}H_{\mu\nu}\right)}\ .
\end{array}\end{equation}
Here $^{(1)}H_{\mu\nu}$ and $^{(3)}H_{\mu\nu}$ are given by
equation (\ref{HIJ}), and
\begin{equation}\begin{array}{rl}\label{PM}
& M_{\rm p}^2(\phi) = M_{\rm p}^2 + 8\pi |\xi|\phi^2 \ ,\\[3mm]
&{\displaystyle M (\phi) = {M_{\rm p} (\phi)\over M_{\rm p} }\
M\ ,}\\[3mm]
&{\displaystyle H_0 (\phi) ={M_{\rm p} (\phi)\over M_{\rm p}
}\ H_0\ .}
\end{array}\end{equation}
During inflation  $(|\dot H| < H^2)$,
the equation of motion for $H$  can be written as follows:
\begin{equation}\label{RSa}
H^2 = {8\pi V(\phi)\over3M_{\rm p}^2(\phi)}
+ {H^4\over H_0^2(\phi)} - {6H^2\dot H\over M^2(\phi)}\ .
\end{equation}
In the limit of $\dot H \to 0$, the equation of motion for the
scalar field reads
\begin{equation}\label{RSb}
3H\dot\phi = 4V(\phi) - V'(\phi) - {3M_{\rm p}^2\over2\pi}\,
H^2\left(1 - {H^2\over H^2_0}\right)\ ,
\end{equation}
On the other hand, equation (\ref{RSa})
in the limit $\dot H \to 0$ has two possible solutions:
\begin{equation} \label{RS}
H^2(\phi)\ = {H_0^2(\phi)\over2} \left(1 \pm \sqrt{1 -
{32\pi V(\phi)\over 3H_0^2(\phi)M_{\rm p}^2(\phi)}}\,\right)\ .
\end{equation}
In the limit of small $\phi$ and $V(\phi)$, the upper branch
corresponds to the Starobinsky inflation with $H^2 = H_0^2$.
Inflation at large $\phi$ is strongly model dependent. The main
qualitative difference with the regime studied in the previous
subsection is the following. If $V(\phi)$ grows at large $\phi$
more slowly than $3H_0^2(\phi)M_{\rm p}^2(\phi)/32\pi$,
then there may be no upper bound for the Hubble constant
$H(\phi)$.

For example, there is no upper bound for  $H(\phi)$ in the
theory  $V(\phi) = {\lambda\over 4} \phi^4$ for $\lambda <
24\pi\xi^2 H_0^2/M_{\rm p}^2$.
In this case, the value of
the inflaton field can increase indefinitely in the regime of
self-repro\-duc\-tion,
\begin{equation}\label{SRF}
\left| 1 - {H^2\over H_0^2} \right| < {H\phi\over2\pi}\ ,
\end{equation}
and we find what we have called runaway solutions, that is, non
stationary probability distributions that move forever towards
large values of the fields \cite{GBLL,JGB}. The rate of inflation
(\ref{RS}) in the limit $8\pi|\xi|\phi \gg M_{\rm p}^2$ at the
upper branch becomes
\begin{equation}\label{RH}
H  \simeq H_0 (\phi) \simeq {H_0\over M_{\rm p}}
(8\pi|\xi|)^{1/2}\,\phi\ ,
\end{equation}
which increases indefinitely with $\phi$. In this case,
self-reproduction occurs for all values of $\phi$ if $8\pi|\xi|
< H_0^2/M_{\rm p}^2$, see eq. (\ref{SRF}).
The rate of inflation at the lower branch also grows without
limit,
\begin{equation}\label{RH1}
H \simeq \sqrt{\lambda\over 12|\xi|}\ \phi \ .
\end{equation}
This regime coincides with the one that we have found at the end
of Section \ref{model2}, see eq. (\ref{PRO}). The reason is very
simple: If the effective Planck mass grows very rapidly with the
growth of $\phi$, then inflation never approaches the Planck
boundary, and the effects associated with the conformal anomaly
always remain small.

\section{\label{CosmConst} Inflation, quantum cosmology and the
cosmological constant problem}

\subsection{Quantum cosmology predictions for the minimal model}
Now we will turn to the discussion of predictions of quantum
cosmology, assuming that one can live in different quantum
states of the universe with different coupling constants. One
should emphasize that this assumption in
its most radical form (all values of coupling constants are
possible) is  extremely speculative, being based on some
particular interpretation of the baby universe theory.   We do not really
know which constants can be considered adjustable, and which
ones are ``true constants''; in what follows we will consider
several different possibilities.

As a working hypothesis, we will assume that the most probable
quantum state of the universe is the state where the total
number of observers of our type can be greater.  This condition
can be rather ambiguous \cite{GBLL}, but we can use it as a
starting point of our investigation of inflationary quantum
cosmology.

The main idea can be illustrated by considering the  simplest
model of the scalar field $\phi$ minimally coupled to gravity
($\xi = 0$) with the effective potential $V(\phi) =
{\lambda\over4}\,\phi^4$. As we already mentioned,  the total
volume of different parts of the universe with a given value of
the scalar field $\phi$ (or with a given density) in this model
grows in time as ${\cal V}(\phi,t) \sim e^{(3-f(\lambda))H_{\max}
t}\ P_p(\phi)$, where $f(\lambda) \to 0$ decreases in the limit
$\lambda \to 0$ \cite{LLM}. It is clear then that the greatest
rate of expansion can be reached in the limit $\lambda \to 0$
\cite{LLM,Vil,Al}.

This is a rather general conclusion.  The   overall rate of
expansion of the universe grows when the effective potential
becomes more and more flat. A similar result was  known  in
chaotic inflation even without taking into account the universe
self-reproduction: The total degree of inflation there was
proportional to $\exp({c/\sqrt\lambda})$, where $c \sim 1$\,
\cite{MyBook}.  Thus, the size the universe after inflation
becomes exponentially large for small $\lambda$. Moreover, it
was known that if one has several scalar fields $\phi_i$ with
coupling constants $\lambda_i$, the last stages of inflation are
typically driven by the field $\phi_i$ with the smallest
$\lambda_i$. This helped, to some extent, to understand why
coupling constants of the inflaton field are so small: They may
be large, but the structure of the part of the observable part
of the universe was formed at the last stage of inflation, which
was driven by the field with the smallest coupling constant
$\lambda_i$ \cite{MyBook}. However, the   results   obtained in
\cite{LLM} are much stronger, and, being interpreted in a
certain way,  they can be even dangerous.

Indeed, let us take the baby universe theory seriously and
assume that we can actually  compare different universes with
different coupling constants. The total number of observers of
our type which may appear  for  a given set of coupling
constants $\lambda_i$ at a given time $t$ is presumably given by
the following symbolic equation:
\begin{equation}\label{life1}
{\cal N}(\lambda_i,t) \sim  \int d \rho_0\, d\rho \,
P_{\rm creat}(\rho_0,\lambda_i)\, P_{\rm life}(\rho,\lambda_i)\,
\int_0^{t} {\cal V}(\rho,t',\rho_0,\lambda_i)\  dt'  \ .
\end{equation}
Here $P_{\rm creat}(\rho,\lambda_i)$ is the probability of
creation of an (inflationary) universe with initial density
$\rho_0 \sim V(\phi_0)$ in the theory with coupling constants
$\lambda_i$, $P_{\rm life}(\rho,\lambda_i)$ is the probability
for life of our type to appear in a unit volume of such a
universe at density $\rho$. This equation  can be written more
accurately, but  the main point we are going to make will not
depend on many details.

Namely, suppose that the probability distribution ${\cal
V}(\rho,t,\rho_0,\lambda_i)$ is stationary, ${\cal
V}(\rho,t,\rho_0,\lambda_i)  \sim e^{\alpha(\lambda_i)\, t}\
P_p(\rho)$, as in eqs. (\ref{SOL}), (\ref{SOL2b}) \cite{time}. It
is important  that this solution {\it does not} depend on the
initial condition $\rho_0$. Therefore one can take an integral
over $\rho_0$, absorbing all information about the probability
distribution $P_{\rm creat}(\rho_0,\lambda_i)$ into some
function $P_{\rm creat}(\lambda_i)$. This yields
\begin{equation}\label{life2}
{\cal N}(\lambda_i,t) \sim  e^{\alpha(\lambda_i)\, t} \
P_{\rm creat}(\lambda_i)\, \alpha^{-1}(\lambda_i) \int d\rho \,
P_{\rm life}(\rho,\lambda_i)\, P_p(\rho)\, \ .
\end{equation}
Typically $P_{\rm creat}(\lambda_i)$ is a smooth  function of
the coupling constants $\lambda_i$. For example,   one can take
for the probability of quantum creation either the square of the
tunneling wave function  $\exp\Bigl(-{3M_{\rm p}^4\over
8\,\rho_0}\Bigr)$ \cite{Creation,Creation2}, or the square of
the Hartle-Hawking wave function  $\exp\Bigl({3M_{\rm p}^4\over
8\,\rho_0}\Bigr)$ \cite{HH} (which, in our opinion, does not
describe quantum creation of the universe \cite{MyBook}).
Independently of this choice, after the integration over
$\rho_0$ one obtains an irrelevant normalization constant
 $P_{\rm creat}$, which
does not depend on $\lambda_i$. The  functions
$\alpha(\lambda_i)$ and $P_{\rm life}(\rho,\lambda_i)$ entering
equation (\ref{life2}) are also not expected to exhibit any
singular behavior with respect to $\lambda_i$. The only function
which strongly depends on $\lambda_i$ in this equation is
$e^{\alpha(\lambda_i)\, t}$.

For example, if the upper boundary for inflation  in the  theory
${\lambda\over 4} \phi^4$ coincides with $V(\phi) = M_{\rm
p}^4$, one has $e^{\alpha(\lambda_i)\, t} \sim
e^{(3-f(\lambda))H_{\max} t} \sim e^{(3-f(\lambda))
\sqrt{8\pi\over 3}M_{\rm p} t}$. It is clear from eq.
(\ref{life2}) that most of the observers of our type should live
at an indefinitely large time interval from the big bang, i.e.
at $t \to \infty$ \cite{LLM}. (The conditions for appearance of life
in each particular domain of the universe at a given density $\rho$
does not depend on time in the limit $t \to \infty$, whereas the
number of such domains grows as $e^{\alpha(\lambda_i)\, t}$.)
But in this limit  the relative
number of observers living in the universes with  a nonvanishing
$\lambda$  becomes suppressed by a factor $e^{-f(\lambda)
\sqrt{8\pi\over 3}M_{\rm p}\, t}$  as compared with the number
of observers living in the universes with $\lambda \to 0$.

Thus the factor $e^{-f(\lambda)\sqrt{8\pi\over 3}M_{\rm p}\, t}$
always wins over all anthropic considerations \cite{GBLL}.  (A
similar argument holds in more traditional versions of the
baby universe theory \cite{RubShap}.) Meanwhile, the
probability of existence of life of our type becomes strongly
suppressed in a universe with small $\lambda$. For example, the
typical amplitude of density perturbations produced during
inflation is proportional to $10^2 \sqrt\lambda$. Therefore, one
could argue that in the limit $\lambda \to 0$ there will be no
galaxies, and no people to live there. The conclusion that
quantum cosmology picks up flat potentials was interpreted in
\cite{Vil,Al} as suggesting that inflation does not produce
density perturbations (these perturbations decrease as $10^2
\sqrt\lambda$ in the limit $\lambda \to 0$), and one should use
topological defects in order to account for galaxy formation.
However, in the limit of absolutely  flat potentials there will
be neither inflationary density perturbations nor topological
defects. Typically, superheavy topological defects are produced
after inflation only if the potential $V(\phi)$ is very curved;
in fact, it is very difficult to produce such defects in the
context of inflationary cosmology. In certain cases these
difficulties can be somewhat weakened, and strings can be
produced even at low energy density in some theories with very
flat potentials \cite{Strings}--\cite{Liddle}. On the other
hand, in the same class of theories it is also possible to
obtain ${\delta \rho\over \rho} \sim  5\times10^{-5}$
 (the result following from the COBE data in the normalization
of ${\delta\rho\over \rho}$ used in \cite{MyBook}) even for extremely
 flat potentials at a very small energy density $V(\phi)$
\cite{Hybrid,Liddle}. Thus, the use of topological defects
produced after inflation may not have any obvious advantages
over the standard inflationary mechanism of generation of
density perturbations. The only constraint on the applicability
of each of these mechanisms is the reheating constraint: If the
energy density at the end of inflation is too low, then particle
production will be exponentially suppressed, and there will be
no observers to enjoy life in such a universe.

This is a typical anthropic constraint and, as we have already
mentioned, it is not strong enough to win over the factor
$e^{-f(\lambda)\sqrt{8\pi\over 3}M_{\rm p}\, t}$. For example, the
expression ${\delta \rho\over \rho} \sim 10^{2}\sqrt\lambda$ is
only statistically correct, which means that in the eternally
self-reproducing inflationary universe there will always appear
 exceptional domains where ${\delta \rho\over \rho} \sim
5\times10^{-5}$ even for $\lambda \ll 10^{-13}$.   Such parts of the
universe may be
extremely rare, but eventually (because of the factor
$e^{-f(\lambda)\sqrt{8\pi\over 3}M_{\rm p}\, t}$) their total volume will
become
much greater that
the volume of   more regular domains with  $\lambda \sim 10^{-13}$
and ${\delta \rho\over \rho} \sim 10^{2}\sqrt\lambda$. The same arguments may
apply to the possibility of large fluctuations with an unusually high baryon
density.
This suggests
that according to our scenario most of the observers should live
in those parts of the universe where conditions for existence of
 life appear as a result of an extremely improbable
fluctuation. The total number of observers living in such
domains should be large  only because of the indefinitely large
time of existence and exponential growth of a self-reproducing
inflationary universe. One could expect that most observers of
our type should live in terrible and irregular conditions,
perhaps on the verge of being extinct. Many would argue that
this conclusion contradicts observational data, even though some
pessimists would agree with this conclusion and use it as an
advanced version of the doomsday prediction.

Since this conclusion  certainly looks unpleasant,
let us see whether we have real   reasons to  worry.

1.There is a simple formal reason to have doubts about the
results of the approach developed above. If the universe enters
the stage of eternal self-reproduction, then after a
sufficiently large time there is no difference between the
universes formed at  two different moments. That is why we
introduced the integration over time $t$ in eq. (\ref{life1}).
However, we also introduced a cutoff in this integral at some
time $t$. The only reason of doing so is that otherwise the
integral diverges and we get infinitely large number of
observers for any given $\lambda$. This not well motivated
introduction of the upper bound in the integral in (\ref{life1})
is the main reason why  the probability distribution $P_p$ does
depend on the choice of time parametrization \cite{LLM,GBLL}.
Therefore, one may argue that the versions of the theory which
allow self-reproduction of the universe and, consequently,
infinitely large number of observers, are preferable as compared
with the versions without self-reproduction, where the number of
observers may be finite. However, if there are many
branches of the universe which allow  self-reproduction,
their comparison is ambiguous
since it involves comparison of infinities, see Introduction.

Unfortunately, this simple argument does not sound entirely
convincing in our case. Indeed, as we already mentioned in
Section \ref{minimal},  even the choice of a radically different
time parametrization $\tau$, where time is measured not by clocks
but by rulers ( $\tau$ is the logarithm of the local expansion
of the universe) does not change our conclusion: The total
volume of all inflationary domains with a given $\phi$ (or with
a given $\rho$) grows at a much greater rate in the limit
$\lambda \to 0$.  Nevertheless, it remains not quite clear
whether   one has any physical reason to compare {\it different
universes  at the same time t}.

2. Our  investigation  was based on the assumption that
different universes with different values of the coupling
constant $\lambda$ may actually exist.  Moreover, we assumed
that the coupling constant $\lambda$ may take  all possible
values, including zero. In the context of the baby universe
theory these assumptions look reasonable, but the baby universe
paradigm may be wrong, or it may have limited validity, being
applicable to the vacuum energy of the universe (cosmological
constant), but not to other parameters. Also, the problem
disappears if $\lambda$   can take only a discrete number of
values, not including zero.

3. Our estimate gives us the total number of observers of our
type for different values of $\lambda$. But is it correct to say
that we consider observers {\it of our type} if they belong to
the universe with different coupling constants? Shall we perhaps
compare sheep to sheep and wolves to wolves? Does it make any
sense to say that it was improbable for the   authors to be born
in Spain and in Russia, because  they could have born in China or
India where the total population is much larger?

4. In our discussion we assumed that the number of observers of
our type is directly proportional to the volume of the universe.
But one cannot get any crop even from a very large field without
having seeds first. The idea that life appears automatically
once there is enough space to be populated may be too primitive.
 More generally, in the present paper we are making an attempt to
describe emergence of life solely in terms of physics.  It is
certainly a most economical approach, but this approach may not
be correct, especially if consciousness has its own
degrees of freedom \cite{MyBook}.

Unfortunately, we are not in a position to discuss  these issues
in the present paper. Quantum cosmology is developed by trial
and error, and we are not pretending to have final answers to
all these questions. Instead of that we will try to develop our
scheme somewhat further, in an attempt to see whether the
conclusions we have reached in this section are general or not.

\subsection{Quantum cosmology in more complicated models of
inflation} The model which we considered in the previous section
has only one dimensionless parameter, $\lambda$. Meanwhile, in
the model (\ref{S2})
there appears another parameter, $\xi$.  We do not really know
whether or not one is allowed to vary each of these two
parameters in the context of the baby universe theory.  Let us
assume for a moment that the parameter $\xi$ is fixed and
positive, as in the first model considered is Section
\ref{model2}. We will also assume that $\xi \ll 1/6$, since
otherwise we do not have inflation in this model, which makes
the total volume small and finite.  In the  theories with  $\xi
\ll 1/6$ there are several different possibilities discussed in
Section  \ref{model2}.

\begin{itemize}
\item [1.] ${\displaystyle {\sqrt\lambda < 64\pi  \xi^2}}$  . \ \
There is inflation but no self-reproduction. In this case
stationarity of $P_p$ is impossible, and the growth of  volume
of inflationary domains is finite.
\item [2.] ${\displaystyle {64\pi  \xi^2<\sqrt\lambda < 16\pi \xi}
}$  . \ \ There is inflation and
self-reproduction.
\item [3.] ${\displaystyle \sqrt\lambda > 16\pi \xi }$  .\ \
In this regime we recover the usual results of chaotic inflation
with a quartic potential.

\end{itemize}

Thus, it is unfavorable (from the point of view of increasing
the number of observers of our type ${\cal N}(\lambda,t)$) to
have $\sqrt\lambda < 64\pi \xi^2$. On the other hand, for  $
\sqrt\lambda > 16\pi \xi$ the number ${\cal N}(\lambda,t)$ grows
with the decrease of $\lambda$, as we have shown in the previous
subsection. This pushes the coupling constant $\lambda$ into the
region
\begin{equation}\label{xi}
64\pi \xi^2<\sqrt\lambda < 16\pi \xi \ .
\end{equation}
In other words, if the coupling constant $\xi$ is fixed,   the
maximum of ${\cal N}(\lambda,t)$ appears not at $\lambda = 0$,
but somewhere in the interval (\ref{xi}).
This corresponds to  density perturbations in the interval
$2\cdot 10^4  \xi^2 < {\delta\rho\over\rho} < 5\cdot 10^3 \xi$.
Thus, instead of the problem of
vanishing  ${\delta\rho\over\rho}$ we have the usual
problem of requiring the coupling constants to be very small, in this
case related to the coupling constant $\xi$.

On the other hand, if one can vary both $\lambda$ and $\xi$, the
results appear to be quite different. In this case the greatest
number of observers will live in the universes with $\xi < 0$,
$\sqrt\lambda < 16\pi |\xi|$. Indeed, as follows from eq.
(\ref{SST}), there exist some domains with a finite volume, in
which the scalar field $\phi$ reaches infinitely large values
within finite time $t = {24\pi|\xi|\over \lambda \phi_0}$. After
that time the rate of growth of the total volume of the universe
containing indefinitely large field $\phi$ becomes infinite, for
any value of $\lambda$. It takes some more time (because of
quantum jumps and classical rolling) before the rate of growth
of volume of the universe with small $\phi$ (and $\rho$) also
becomes infinite. It is important, however, that the rate of
expansion of domains containing a finite field $\phi$ (or finite
density $\rho$) becomes infinitely large within some finite
time.  It hardly makes any sense to compare the number of
observers in different universes  if this number is
infinite in each universe even at a finite time $t$. This is a
must stronger manifestation of the same ambiguity which we
encountered before, when the integral in (\ref{life1}) became
infinite in the limit $t \to \infty$.

Our conclusions will not change much if we add the terms $R^2$
to our model.  As follows from eq. (\ref{DIM2}), the overall
rate of expansion of the whole universe grows when the curvature
of $V(\phi)$ (i.e. the mass $m$) decreases. On the other hand,
if one can vary both $\lambda$ and $\xi$ in the theory
${\lambda\over 4} \phi^4$, then for negative $\xi$, and $\lambda
< 24\pi\xi^2 H_0^2/M_{\rm p}^2$
one obtains runaway solutions, and the rate of
expansion becomes infinitely large within a finite time.

Let us summarize our results. The theory ${\lambda\over
4}\phi^4$ with $\xi= 0$ is a particular version of the more
general class of models with arbitrary $\xi$. In fact, even if
one starts with the model with $\xi = 0$ for $\phi = 0$ and $R =
0$, one almost inevitably obtains an effective coupling constant
$\xi \not = 0$, which depends logarithmically  on $\phi$ and $R$,
as a result of quantum corrections \cite{114B}. If one varies
$\lambda$ for a given $\xi > 0$, one does not obtain the
dangerous result that ${\cal N}(\lambda,t)$ has a maximum at
$\lambda = 0$. Instead one can find a maximum of   ${\cal
N}(\lambda,t)$ somewhere in the interval $64\pi \xi^2<
\sqrt\lambda < 16\pi \xi$. On the other hand, if one can vary
both $\lambda$ and $\xi$, then the only conditions which one can
get are $\xi < 0$ and $\sqrt\lambda < 16\pi |\xi|$. Under these
conditions the total volume of the universe (and the total
number of observers which will occupy this volume later) becomes
infinite within a finite time $t$, which makes any further
analysis ambiguous. Therefore at present we do not think that
one should worry too much about the conclusion that quantum
cosmology prefers vanishing effective potentials $V(\phi)$, even
though this issue deserves further consideration.

In addition, we should emphasize  again that  all results
obtained in this section have been derived under the very speculative
assumption that one can compare   {\it different universes} at the same time.
If one would compare different exponentially large {\it parts of the universe}
with different laws of low-energy physics, there would be no difference in the
rate of growth  $e^{\alpha(\lambda_i) t}$ of these parts \cite{LLM,GBLL}, and
our conclusions would be quite different, see Discussion.

\subsection{Extended Starobinsky model and the cosmological
constant problem}
Even though there are many potential problems
associated with the approach developed in this paper, it certainly allows us to
look at
the old problems of quantum cosmology from a different
perspective. Let us try to apply our methods to the problem of
the cosmological constant.

The main lesson we have learned from our previous investigation
is that the total number of observers is almost entirely
controlled by the factor $e^{d H_{\rm max}t}$. If one considers
the usual inflationary models where inflation is driven by the
scalar field $\phi$, then one can expect that adding a vacuum
energy $V_0$ to the effective potential will only increase $d
H_{\rm max}$.  For example, if the upper boundary for inflation
is determined by the condition $V(\phi) = M_{\rm p}^4$, then $H_{\rm
max} = \sqrt{8\pi\over 3}M_{\rm p}$, independently of $V_0$. A
similar result is valid if there is no inflation in Starobinsky
model (if e.g. $H_0 \ll M$),
and the upper bound for the scalar-field-driven inflation is
given by $H_0/\sqrt 2$, independently of $V_0$. Then the only
factor which depends on $V_0$ is the fractal dimension $d$. One
can easily understand, for example, that with an increase of $V_0$ the
potential ${m^2\over 2}\phi^2 +V_0$ near its upper bound
becomes more flat, which typically increases $d$. We have
verified this conjecture by changing $V_0$ and finding $d$
numerically.  Thus one may guess that quantum cosmology pushes
the cosmological constant $\Lambda \equiv {8\pi V_0\over M_{\rm
p}^2}$ towards its largest possible values. This would be a very
undesirable conclusion, especially since the anthropic principle
allows a positive vacuum energy $V_0$ to be as large as $10^{-27} g
\cdot cm^{-3}$, which is two orders of magnitude greater than
the present observational constraints on $V_0$.

However, this conclusion is not general. Indeed, let us consider
Starobinsky model together with the scalar field with the
effective potential $V(\phi) + V_0$, where $V_0$ is some
constant, and $V(\phi)$, as before, vanishes in its minimum, e.g. $ V(\phi) =
{m^2\over 2}\phi^2$.  We
will assume that $H_0 \ll M_{\rm p}$, and ${H_0^2\over M_{\rm
p}} \sqrt {3\over \pi} \ll M < 3 H_0$. As we have shown in the Section
\ref{model3}, the maximal
rate of expansion $H_{\rm max} \simeq H_0 +
\Delta H \simeq H_0$
is achieved at top of the Starobinsky branch, for the smallest
value of the effective potential $V(\phi)=0$, and the
self-reproduction of the universe occurs in a very narrow region
near $H_0$, $\Delta H \ll H_0$ (\ref{IAS}). However, in the
present case the analogue of  eq. (\ref{RS3a}) at small  $ \phi$
yields
\begin{equation}\label{CCS}
H^2 = {H_0^2\over2}\,\left(1 \pm \sqrt{1 -
{4\Lambda\over3H_0^2}}\right)\ ,
\end{equation}
where $\Lambda = {8\pi\over M_{\rm p}^2} V_0$ is the
cosmological constant. Thus, adding the cosmological constant
$\Lambda > 0$  diminishes the maximal value of the Hubble
constant $H$. In the limit $\Lambda \ll H_0^2$ one obtains
\begin{equation}\label{DSP}
H_{\rm max} \simeq H_0\left(1 - {\Lambda\over6H_0^2}\right)\  .
\end{equation}
Adding a cosmological constant changes the
fractal dimension $d$ as well. The corresponding changes can be
approximately described by substituting  $H_{\rm max}$  for  $H_0$
in (\ref{DIM2}):
\begin{equation}\label{FrDim}
 d(\Lambda) =\ 3 - {M^2\over3H_0^2} - {M^2\Lambda\over9H_0^4}  - {m\over
2\sqrt\pi M_{\rm p}}\ .
\end{equation}
This gives
\begin{equation}\label{COSM}
e^{d H_{\rm max}t} \approx  \exp\left[d(\Lambda)\, H_0\left(1 -
{\Lambda\over6H_0^2}\right)\,t\right]\ .
\end{equation}
Note that both factors in the exponent decrease with a growth
of $\Lambda$.  In the $\tau$-parametrization of time the last
factor disappears, being absorbed into the definition of $\tau$
\cite{LLM}, and the resulting exponential factor acquires the
following form:
\begin{equation}\label{COSMtau}
e^{d(\Lambda)\,\tau} =  \exp\left[ \left(3 - {M^2\over3H_0^2} -
{M^2\Lambda\over9H_0^4}   - {m\over 2\sqrt\pi M_{\rm p}}\right)\tau\right]\ .
\end{equation}
Therefore, the exponential factor decreases with increasing
$\Lambda$ for either choice of the time para\-metri\-zation.  This
suggests that out of all possible universes with $\Lambda \geq
0$ it is most probable to live in those universes with $\Lambda
\simeq 0$. Perhaps this is the reason why we live in a
universe (or in a part of the universe) with a vanishingly small
value of the cosmological constant!

Note that both in this case and in the case of the coupling
constant $\lambda$ we are speaking about probability
distributions that become infinitely sharp in the limit $t \to
\infty$. It strongly resembles the infinite sharpness of the
distribution of probability to find a universe with a given
cosmological constant in the context of the baby universe
theory, $P(\Lambda) \sim \exp \Bigl(\exp {3\pi M^2_P\over
\Lambda}\Bigl)$ \cite{Coleman}. The existence of such a peak in the
baby universe theory was extremely counterintuitive.  Indeed, if
our universe had lived for only $10^{10}$ yrs, it could not
``know'' the value of its energy density with infinitely high
precision. In our case the explanation of the infinite sharpness
of the probability distribution is very simple: The universe
lives for infinitely long time, and even a very small deviation
from $\Lambda = 0$ eventually makes a lot of difference.

A note of caution is needed here before one gets too excited.
First of all, we have obtained our results in the context of a
particular inflationary theory, which may or may not be correct.
Our conclusions would be different if one is allowed to vary not
only the cosmological constant but other coupling constants as
well. Even more immediate problem arises if one considers the
possibility to have a negative cosmological constant, since in
this case our exponential factor becomes even greater.  This is
similar to the problem of the negative cosmological constant in
the baby universe theory
\cite{Rub}.

An obvious way out of this difficulty is to note that the
universe with $V_0 \ll -10^{-29} g \cdot cm^{-3}$ would collapse
within the time smaller than $10^{10}$ yrs, and nobody would
discuss the cosmological constant problem in such a universe.  Unfortunately,
as we
already mentioned, anthropic considerations typically are not
strong enough to fight against indefinitely growing or
decreasing exponents.  However, anthropic considerations could
be quite sufficient if we were able to find some natural cutoff
in our integrals (\ref{life1}) at very large $t$. Taking into
account all our doubts concerning the measure of integration,
this possibility is not inconceivable.

Another possibility is that a negative cosmological constant
is forbidden by some law of nature. This is known to be the case
in globally supersymmetric theories, where the cosmological
constant can only be positive or zero. In locally supersymmetric
theories this property can be violated. Still there is a chance
that in future theories the problem of a negative cosmological
constant will be less urgent.

Finally, it is quite possible that the problem of a negative
cosmological constant should be addressed at a somewhat more
advanced level. If there is any analogy between our approach and
the baby universe theory, this analogy suggests that perhaps we
still did not make ``exponentiation of the exponent'', we still
did not take into account non-local interactions of
exponentially expanding universes with each other. This
interaction may be less efficient if the universes must
disappear soon after being created, which is the case if the
cosmological constant is negative. A possible solution of the
problem of negative cosmological constant in the context of the
inflationary baby universe theory was envisaged in ref.
\cite{Mezh}.

\section{\label{Discussion} Discussion}
This paper consists of two main parts. In the first part of the
paper, Sections 2--4, we have studied several different regimes
which are possible in inflationary cosmology with an account
taken of the process of self-reproduction of inflationary
domains. It appears that by changing the coupling constants in a
simple class of inflationary models one can  go from the models
where inflation is possible, but there is no self-reproduction
of the universe, to the models where the universe is
self-reproducing, and it  can be described by a stationary
probability distribution $P_p(\phi)$. Some additional
modifications lead  to models where self-reproduction of the
universe is so active that the corresponding probability
distribution within finite time moves towards infinitely large
values of the inflaton field $\phi$.  This classification of
possible inflationary regimes, as well as the investigation of
self-reproduction in the models of a scalar field  nonminimally
coupled to gravity and in Starobinsky model, can be of some
interest independently of a more speculative discussion
contained in Section \ref{CosmConst}.

The investigation performed in Section \ref{CosmConst} is based
on the assumption that   coupling constants can take different
values   in different quantum states of the universe (which we
call different  ``universes'').    The basic assumption which we
are making is that we are   typical observers, and therefore we
live in those universes where most other observers live. Thus,
by finding out the values of coupling constants in the universes
occupied by most of the observers, we may find an
``explanation'' of the values of the coupling constants in our
own universe.

This approach is very ambiguous, even though we understand it
much better than the Euclidean approach to the baby universe
theory. It is amazing that in some cases these two approaches
give similar results. One of these results seems especially
interesting.  If one considers self-reproduction of the universe
in the context of the Starobinsky model and assumes that the
cosmological constant $\Lambda$ may take different non-negative
values in different parts of the universe, then our results
imply that a typical observer  should live in a state with
$\Lambda = 0$.  Of course, one should not consider this result
as a   solution of the cosmological constant problem until we
understand why the cosmological constant cannot be negative. We
clearly realize how far we are from any final and rigorously
proven conclusions.

In the beginning of our paper we
have mentioned that the possibility of having different coupling
constants appears even without any recourse to the baby universe
theory.  Indeed,  during the process of self-reproduction
of inflationary domains the Universe becomes divided into
exponentially large regions where all possible laws of the
low-energy physics compatible with inflation can be realized
\cite{MyBook}. Since these parts of the universe are
exponentially large and  causally disconnected, for all
practical purposes one may consider them as separate universes.
Thus one could expect that all results of the investigation
performed in Section \ref{CosmConst} should be valid for the
distribution of probability to live in a part of the universe
with  given values of coupling constants. This would make our
discussion much less speculative.

 We have used this approach in \cite{LLM,GBLL}. However,   the
results obtained in \cite{LLM,GBLL}  differ considerably from the
results of our investigation of the baby universe theory. The
reason is that in all parts of the universe which can be
produced from a single inflationary domain by the process of
classical motion and quantum diffusion (or tunneling), the
exponential factor $e^{\alpha t}$\, in the stationary
distribution (\ref{SOL}) is universal. It does not depend on the
value of the effective cosmological constant (vacuum energy) in each
particular minimum of the effective potential, on the curvature
of the effective potential near each of its minima, etc. All
phases which can exist in the theory and which can transform to
each other due to classical motion and quantum jumps appear to
be in a kind of  ``thermal equilibrium''  with the same
``temperature'' $\alpha$.  Only those parts of the phase space
of the theory  which evolve absolutely independently of  each other
can have different values of $\alpha$
\cite{ColWein}. Therefore in the theories where the probability
distribution is stationary, the most important tool for
comparison of different branches of inflationary universe is not
the overall factor $e^{\alpha t}$, which we have studied in the
present paper, but the normalized
probability distribution $P_p$. This distribution does not have
any singularities encountered in our treatment of the baby
universe model.  Consequently, this approach  is not expected to
lead to any troubles with too flat effective potentials. Since the
probability distribution $P_p$ is not singular, one
can use it in combination with the anthropic principle in order to
explain the values of  effective coupling constants in our part
of the universe \cite{LLM,GBLL}.

One the other hand, the  theory of inflationary baby universes
can be more powerful in solving  the cosmological constant
problem.  That is why we have mentioned both possibilities in
the present paper.  The main reason why we decided to discuss
here the results obtained by both methods despite all
uncertainties involved can be explained as follows. For many
years the general attitude towards quantum cosmology was rather
sceptical. Even some authors of quantum cosmology believed that
this theory, being very important for investigation of creation
of the universe,   does not have any testable observational
consequences. However, our investigation suggests
that within the context of quantum cosmology there may exist a
rather strong relation between the values
of coupling constants, the structure of the universe at
ultimately large distances, and the properties of interactions
at nearly Planckian scales. By testing one part of the picture
we may get some nontrivial information about its other parts.
This may give us a possibility to test the basic principles of
quantum cosmology as well.

\section{Acknowledgements}

It is a pleasure to thank  L.A. Kofman, A. Mezhlumian  and A.A.
Starobinsky for valuable discussions at different stages of this
investigation. The work of A.L. was supported in part by NSF
grant PHY-8612280. The work of J.G.-B.  was supported  by  a
Spanish Government MEC-FPI postdoctoral grant.


\begin{thebibliography}{999}
\bibitem{time} Note, that the time of establishing this stationary
regime is extremely short for the part of the distribution at
large $\rho$ \cite{LLM}, but it becomes increasingly larger for
small $\rho$.
\bibitem{diff}  Here we somewhat differ from ref. \cite{Vil},
where it was supposed that eq. (\ref{SOL2}) is valid only at
large $\phi$, where quantum fluctuations are very large.
\bibitem{ColWein} This possibility may appear e.g. in the
Coleman-Weinberg theory, where inflation is possible both at
$\phi = 0$ and at $\phi > M_{\rm p}$.  These two inflationary
branches are divided by some interval where inflation is
impossible. Therefore if one neglects  extremely improbable
large jumps from one branch  to another, the total volume
corresponding to each branch will grow at a different rate (in
time $t$), and the volume corresponding to the branch with $\phi
> M_{\rm p}$ eventually will dominate \cite{LLM}.
\end{thebibliography}
\end{document}